# A THREE-CLASS ROC FOR EVALUATING DOUBLETALK DETECTORS IN ACOUSTIC ECHO CANCELLATION[*]

[1]*Jianming Liu*, [2]*Weiqian Liang*, [2]*Runsheng Liu*

[1]Institute of Microelectronics of Tsinghua University, Beijing, 100084, China
[2]Department of Electronic Engineering, Tsinghua University, Beijing, 100084, China

**ABSTRACT**

Doubletalk detector (DTD) is essential to keep adaptive filter from diverging in the presence of near-end speech in acoustic echo cancellation (AEC), and there was a receiver operating characteristic (ROC) to characterize DTD performance. However, the traditional ROC for evaluating DTD used a static time-invariant room acoustic impulse response and could not evaluate DTDs which distinguish echo path change from doubletalk. We solve these problems by extending the traditional binary detection ROC to three-class, and simulations show the efficiency of the proposed method.

## 1. INTRODUCTION

The problem of acoustic echo cancellation is usually done by modeling the echo path impulse response with an adaptive filter and subtracting the echo from the microphone output signal. Three situations exist in echo cancellation: single-talk, doubletalk and echo path change. The adaptive filter should update its coefficients during the single-talk periods, even more rapidly during the echo path change periods, and freeze its coefficients during the doubletalk periods.[1]

Different doubletalk detectors have been proposed, and Fig.1 shows the overview of a generic ideal doubletalk detector. Generally, it is handled in the following way:

In the first stage, a detection statistic $\Phi$ is formed using available signals, e.g., far-end speech, microphone signal, error signal of adaptive filter, etc., and the estimated filter coefficients. $\phi = (\Phi > T1)$ means the detection statistic $\Phi$ is compared to a preset threshold $T1$, when $\Phi > T1$, $\phi = 1$ and doubletalk or echo path change is declared. If $\Phi \leq T1$, $\phi = 0$ and far-end is declared.

[*]This work was supported by the Hi-Tech Research and Development (863) Program of China under Grant No.2008AA010700. Corresponding author: Weiqian Liang, Assistant Professor, Email: lwq@tsinghua.edu.cn.

Traditionally we call this stage as doubletalk detection (DTD). An accurate distinction between the doubletalk and echo path change is made in the second stage named as echo path change detection (EPCD). The second detection statistic E is calculated, and compared to another preset threshold $T2$. Doubletalk is declared if $E \leq T2$, $\varepsilon = 1$, or echo path change when $E > T2$, $\varepsilon = 0$.

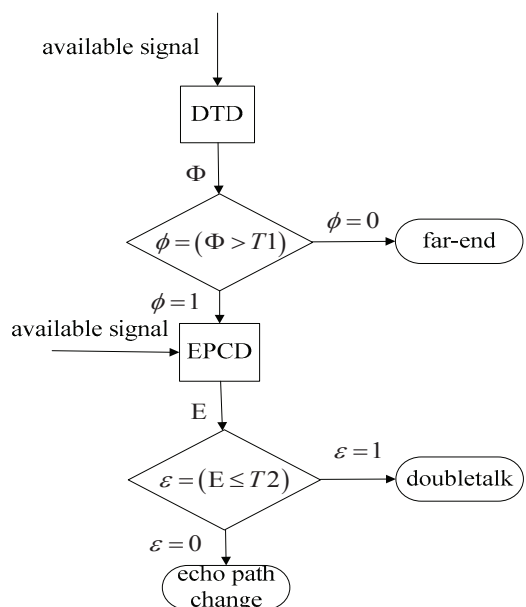

**Fig.1** Block diagram of generic ideal doubletalk detector.

However, most of the current DTD algorithms [1] [2] [5] do not distinguish echo path changes from doubletalk.

There was a binary detection ROC method to objectively select the threshold and compare different DTD algorithms [1], however, it did not consider echo path changes, which might be misleading in the presence of echo path changes. Therefore, we extend the binary ROC for DTD to three-class in order to evaluate these DTDs objectively in the presence of echo path changes.

This paper is organized as follows: Section 2 describes the traditional binary ROC. The extended three-class ROC analysis is proposed in Section 3. Section 4 presents the simulation results. Conclusions are drawn in Section 5.

## 2. TRADITIONAL BINARY DETECTION ROC

The traditional DTD evaluation method viewed DTD as a binary detection problem with special considerations to the AEC application [1]. The binary ROC for evaluating DTD was as follows.

*Probability of false alarm* ( $P_f$ ): probability of declaring doubletalk when doubletalk is not present.

*Probability of miss* ( $P_m$ ): probability of declaring no doubletalk when doubletalk is present.

The probability of false alarm at each threshold point is calculated with no near-end speech as

$$P_f = \frac{\sum x \cdot \phi}{N} \quad (1)$$

where $\phi$ is the DTD output, $x$ is the voice activity detector (VAD) output of far-end speech, and $N$ is the length of entire far-end speech signal.

Then the near-end speech is applied, and the miss probability $P_m$ is calculated as

$$P_m = 1 - \frac{\sum x \cdot v \cdot \phi}{\sum x \cdot v} \quad (2)$$

where $v$ is the VAD output of near-end signal.

The evaluation is based on taking the probability of miss $P_m$ as a function of near-end to far-end speech ratio (NFR) under a given $P_f$ [1]. However, when there are echo path changes, it is difficult to make a fair comparison between DTDs, and [3] gives two reasons for this: as the SNR is computed as an average over the entire data set, it will vary when the damping in the acoustic echo path varies. Furthermore, the threshold value is always generated using the post-change impulse response. If there is a change in the path damping, and thus in the SNR and NFR, the selected threshold does not necessarily appropriate for the pre-change data. We solve this problem by extending to three-class ROC in the next section.

## 3. THE EXTENDED THREE-CLASS ROC

Let us denote the VAD output of far-end signal, echo signal, near-end signal as *x, y, v,* and the output of DTD and EPCD as $\phi$ and $\varepsilon$. We assume that all the echo path changes take place instantaneously, and hold for $T_{hold}$ until the adaptive filter converges to the new echo path again, e.g., for the block normalized least mean square (BNLMS) adaptive filter with 1024 taps and stepsize $\mu = 0.5$ at *fs=8kHz*, $T_{hold}$ is about *1.5s*. This assumption is reasonable because experiments show that the performance of DTD with continuous echo path changes were almost identical to those where the change took place instantaneously, i.e. the duration of the change has almost no impact on the performance of the algorithms [3]. According to the above assumption, we define an echo path change input vector *c*.

Then we derive all the probabilities as listed in Table.1.

**Table.1** Probabilities for each state in DTD

| Status | Probability |
|---|---|
| Declare Far as Far | $P_{ff} = \dfrac{\sum x \cdot \bar{v} \cdot \bar{c} \cdot \bar{\phi}}{\sum x \cdot \bar{v} \cdot \bar{c}}$ |
| Declare Far as Double | $P_{fd} = \dfrac{\sum x \cdot \bar{v} \cdot \bar{c} \cdot \phi \cdot \bar{\varepsilon}}{\sum x \cdot \bar{v} \cdot \bar{c}}$ |
| Declare Far as Change | $P_{fc} = \dfrac{\sum x \cdot \bar{v} \cdot \bar{c} \cdot \phi \cdot \varepsilon}{\sum x \cdot \bar{v} \cdot \bar{c}}$ |
| Declare Double as Far | $P_{df} = \dfrac{\sum x \cdot v \cdot \bar{\phi}}{\sum x \cdot v}$ |
| Declare Double as Double | $P_{dd} = \dfrac{\sum x \cdot v \cdot \phi \cdot \bar{\varepsilon}}{\sum x \cdot v}$ |
| Declare Double as Change | $P_{dc} = \dfrac{\sum x \cdot v \cdot \phi \cdot \varepsilon}{\sum x \cdot v}$ |
| Declare Change as Far | $P_{cf} = \dfrac{\sum x \cdot \bar{v} \cdot c \cdot \bar{\phi}}{\sum (x + \bar{x} \cdot y) \cdot \bar{v} \cdot c}$ |
| Declare Change as Double | $P_{cd} = \dfrac{\sum x \cdot \bar{v} \cdot c \cdot \phi \cdot \bar{\varepsilon}}{\sum (x + \bar{x} \cdot y) \cdot \bar{v} \cdot c}$ |
| Declare Change as Change | $P_{cc} = \dfrac{\sum x \cdot \bar{v} \cdot c \cdot \phi \cdot \varepsilon}{\sum (x + \bar{x} \cdot y) \cdot \bar{v} \cdot c}$ |

Note: the symbol ' · ' denotes logic operator AND, '+' denotes OR, and the line on the vectors means NOT.

We note that:

$$P_{ff} + P_{fd} + P_{fc} = 1 \quad (3)$$

$$P_{df} + P_{dd} + P_{dc} = 1 \quad (4)$$

$$P_{cf} + P_{cd} + P_{cc} = 1 \quad (5)$$

We will show how to use these probabilities to evaluate the DTD with three-class ROC analysis in the following next sub-sections.

### 3.1. Three-class ROC analysis for evaluating DTD performance

If the classifiers for *Q* classes are considered to be points with coordinates given by their *Q(Q-1)* misclassification rates, it is desirable to simultaneously minimize all the misclassification rates, and the optimal classifiers lie on the convex hull of these points [4]. We describe the surface in terms of Pareto optimality and give an evolutionary algorithm for locating the optimal ROC surface and comparing performance of two DTDs'.

Considering the optimal ROC surface as a function of the DTD threshold, $T1$ and $T2$, if all the misclassification rates for one DTD with threshold $\{T1, T2\}$ are no worse than the classification rates for another DTD with threshold $\{T1', T2'\}$, and at least one rate is better, then the DTD with $\{T1, T2\}$ is said to *strictly dominate* that with $\{T1', T2'\}$. Less stringently, $\{T1, T2\}$ *weakly dominates* $\{T1', T2'\}$ if one DTD with threshold $\{T1, T2\}$ are no worse than the classification rates for another DTD with threshold $\{T1', T2'\}$. A set of $\{T1, T2\}$ is said to be *non-dominated* if no member of the set is dominated by any other member [5].

A solution to the minimization problem is thus *Pareto optimal* if it is not dominated by any other feasible solution, and the non-dominated set of all Pareto optimal solutions is known as the *Pareto front*. We describe an evolutionary algorithm to locate the Pareto front of the there-class ROC. In outline, the algorithm maintains a set of archive $F$, whose members are mutually non-dominating, which forms the current approximation to the Pareto front. As the threshold $\{T1, T2\}$ increases, we derive the misclassification rates, and if the new $\{T1, T2\}$ is not dominated by members of the archive $F$, we insert them into $F$, meanwhile any threshold pair in $F$ which are dominated by the new entrant are removed [4]. In order to compare the performance of two DTD algorithms, we locate the Pareto front for each method respectively at first, then mix the two Pareto fronts together and locate the new Pareto front of them.

In the above analysis, we attempt to minimize all the misclassifications from a multi-object perspective, however, specific for the AEC application, the penalty of misclassification is quite different from each other. Actually, a moderately high $P_{fd}$, $P_{fc}$ and $P_{cf}$ is tolerable, and meanwhile the $P_{df}$, $P_{dc}$, $P_{cd}$ characteristic is a meaningful criterion to fairly compare different DTDs. Therefore, we can only focus on the points in certain range of $P_{fd}$, $P_{fc}$ and $P_{cf}$, for example *0.1-0.3*, and investigate the $P_{df}$, $P_{dc}$, $P_{cd}$. Meanwhile, considering the Pareto front is quite not straightforward for selecting threshold and comparing DTD, according to application of AEC, we can simply visualize it by assuming $P_{fd}$, $P_{fc}$, $P_{cf}$ have the same cost, and $P_{df}$, $P_{dc}$, $P_{cd}$ have the same cost. Similar to the approach of taking the probability of miss, $P_m$, as a function of probability of false alarm, $P_f$ in [2], we obtain

$$P_x = \frac{P_{fd} + P_{fc} + P_{cf}}{3} \tag{6}$$

$$P_y = \frac{P_{df} + P_{dc} + P_{cd}}{3} \tag{7}$$

where all the probabilities come from Pareto front, which eliminates the non-optimal points.

Then we could plot the probability we care very much, $P_y$, as a function of probability we care not so much, $P_x$, in two-dimention. Therefore we extend the traditional ROC curve to the situation under echo path changes with three-class ROC.

### 3.2. Relationship with binary ROC

From the definition of probability of false alarm, $P_{false}$, we obtain

$$P_{false} = P_{fd} + P_{cd} \tag{8}$$

when there is no echo path change, i.e. $c = 0$, we set $P_{cd} = 0$ and

$$P_{false} = P_{fd} = \frac{\sum x \cdot \overline{v} \cdot \phi \cdot \varepsilon}{\sum x \cdot \overline{v}} \tag{9}$$

when there is no near-end speech, i.e. $v = 0$, and DTD does not distinguish echo path change from doubletalk, i.e. $\varepsilon = 1$, we get

$$P_{false} = \frac{\sum x \cdot \phi}{\sum x} \tag{10}$$

There is a slight difference in the denominator with (1) in classical ROC though they have the same trend. This is because classical ROC does not consider the effects of pause on probability in the far-end speech, therefore (8) is more reasonable.

From the definition of probability of miss detection, $P_{miss}$, we obtain

$$P_{miss} = P_{df} + P_{dc} = 1 - \frac{\sum x \cdot v \cdot \phi \cdot \varepsilon}{\sum x \cdot v} \tag{11}$$

When DTD does not distinguish echo path change from doubletalk, i.e., $\varepsilon = 1$, we obtain

$$P_{miss} = 1 - \frac{\sum x \cdot v \cdot \phi}{\sum x \cdot v} \tag{12}$$

which is the same as (2) in the classical ROC.

According to the analysis above, we know that the classical ROC could not be used to evaluate the DTD which distinguishes echo path change from doubletalk and is an approximation of the proposed three-class ROC.

## 4. SIMULATION RESULTS AND COMPARISON

We use a recorded digital speech sampled at 8 kHz for far-end and near-end speech and a L=1024-sample room impulse response. We simulate two changes in the echo path: the damping of the impulse responses before the first change is 10 times larger than after the first change, and the damping after the second change is 10 times larger than before the second change [3]. All the changes occur instantaneously and $T_{hold} \approx 1s$. We compare Geigel DTD [5] with cross-correlation DTD [1] using the three-class ROC. Table.2. and Table.3. show part of the Pareto front of Geigel DTD and Cross-correlation DTD.

**Table.2.** Part of Geigel DTD Pareto Front.

| Tl | $P_{fd}$ | $P_{cf}$ | $P_{df}$ | $P_{cd}$ |
|---|---|---|---|---|
| 0.008 | 0.407 | 0.333 | 0.743 | 0.133 |
| 0.024 | 0.196 | 0.417 | 0.857 | 0.050 |
| 0.104 | 0.053 | 0.450 | 0.871 | 0.017 |

**Table.3.** Part of Cross-correlation DTD Pareto Front.

| Tl | $P_{fd}$ | $P_{cf}$ | $P_{df}$ | $P_{cd}$ |
|---|---|---|---|---|
| 0.70 | 0.148 | 0.400 | 0.029 | 0.067 |
| 0.72 | 0.228 | 0.383 | 0.014 | 0.083 |
| 0.74 | 0.365 | 0.317 | 0.000 | 0.150 |

It is noted that $P_{fc} = 0$ and $P_{dc} = 0$ because the two DTDs do not distinguish echo path change from doubletalk.

According to (6) and (7), we plot the probability we care very much, $P_y$, as a function of probability we care not so much, $P_x$, in two-dimention.as in Fig.2, and it is clear that cross-correlation is superior than Geigel under echo path changes.

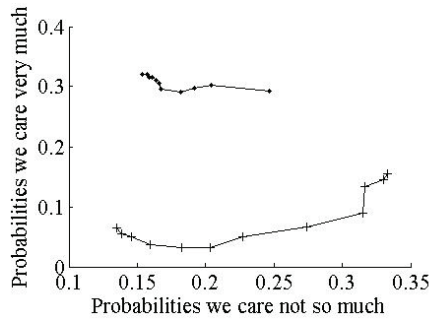

**Fig.2** Comparison between Geigel ( • ) and Cross-correlation ( + ) using three-class ROC.

The performance of DTD under echo path changes depends on the converging speed of the adaptive filter, and we simulate the performance with different $T_{hold}$ as in Fig.3. It shows that the performance degradation with the increase of $T_{hold}$.

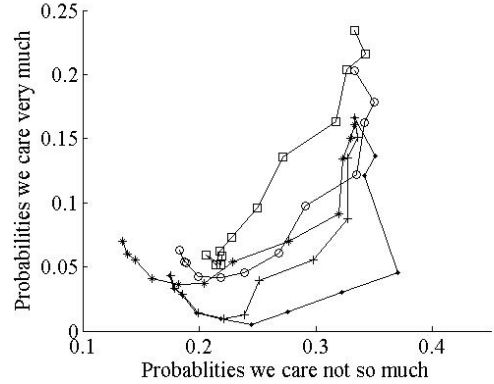

**Fig.3** Cross-correlation DTD with different $T_{hold}$ : 352ms ( • ), 672ms ( + ), 992ms ( ∗ ), 1.3s ( ○ ), 1.6s ( □ ).

## 5. CONCLUSION

In order to evaluate the DTD performance under echo path changes, we extend the classical binary ROC to three-class ROC. We derive the probabilities of misclassification and introduce an evolutionary algorithm to locate the Pareto front with specific consideration of the doubletalk detector in acoustic echo cancellation. Finally, we extend the traditional ROC curve to the situation under echo path changes with three-class ROC. Simulations show that this three-class ROC could evaluate DTD more reasonably in the presence of echo path changes.